\documentclass[prl,aps,twocolumn,showpacs,amsmath,amssymb,floatfix,superscriptaddress,citeautoscript]{revtex4-1}
\usepackage{graphicx}
\usepackage{dcolumn}
\usepackage{bm}
\usepackage{amsmath}
\usepackage{amssymb,bbold}
\usepackage{verbatim}
\usepackage[normal]{subfigure}
\usepackage[bookmarks=false,linkcolor=blue,urlcolor=blue,colorlinks,citecolor=blue]{hyperref}

\newcommand{\beq}    {\begin{equation}}
\newcommand{\enq}    {\end{equation}}
\newcommand{\ceq}[1] {(\ref{#1})}

\newcommand{\eps}    {\epsilon}
\newcommand{\kk}     {{\bf k}}
\newcommand{\qq}     {{\bf q}}
\newcommand{\dd}     {{\bf d}}

\newcommand{\rr}     {{\bf r}}

\newcommand{\ssigma}   {\boldsymbol{\sigma}}
\newcommand{\uimp}     {u_{\rm imp}}
\newcommand{\identity}     {\mathbb{1}}
\newcommand{\ang} {\mbox{\AA}}

\makeatletter

\begin{document}

\title{Impurity-induced states in superconducting heterostructures}

\author{Dong E. Liu}
\affiliation{Station Q, Microsoft Research, Santa Barbara, California 93106-6105, USA}
\author{Enrico Rossi}
\affiliation{Department of Physics, William \& Mary, Williamsburg, VA 23187, USA}
\author{Roman M. Lutchyn}
\affiliation{Station Q, Microsoft Research, Santa Barbara, California 93106-6105, USA}

\date{\today}

\begin{abstract}
Heterostructures allow the realization of electronic states that are difficult to obtain in isolated uniform systems.
Exemplary is the case of quasi-one-dimensional heterostructures formed by a superconductor
and a semiconductor with spin-orbit coupling in which Majorana zero-energy modes can be realized.
We study the effect of a single impurity on the energy spectrum of superconducting heterostructures.
We find that the coupling between the superconductor and the semiconductor
can strongly affect the impurity-induced states and may induce additional subgap bound states
that are not present in isolated uniform superconductors. For the case of quasi-one-dimensional superconductor/semiconductor heterostructures
we obtain the conditions for which the low-energy impurity-induced bound states appear.

\end{abstract}


\maketitle

Composite heterostructures provide an opportunity to realize states with novel and desirable properties
that are different from the individual components. In the last decade, this principle has been implemented very successfully to obtain {\em composite electronic systems}
with novel and unique electronic properties.
For example,
the heterostructures combining a conventional s-wave superconductor (SC) and a semiconductor with strong spin-orbit coupling (SOC) may realize topological superconducting states supporting Majorana zero modes (MZMs)\cite{Hasan2010,Beenakker13a,Alicea12a,Leijnse12,Stanescu13b,Franz'15,DasSarma15,MasatoshiSato2016,lutchynreview2017},
and the preliminary signatures of MZMs were
observed~\cite{Mourik12,Rokhinson12,Deng12,Churchill13,Das12,Finck12,Nadj-Perge14,Deng2014,Higginbotham15,Albrecht16,Deng2016,Zhang16,Zhang17}.
%

The presence of impurities in heterostructures, as in any other condensed matter system, is unavoidable. However, their effect on the electronic states can be quite non-trivial due to the interplay between scattering processes involving different materials. The effect of impurities
in general varies significantly depending on the component of the heterostructure in which they are located.
This fact makes the understanding of impurity effects
in heterostructures non trivial and outside the scope of most previous works focusing on impurity effects in single-component homogeneous systems.


In this work we study the states induced by scalar impurities
in heterostructures involving a SC and a semiconductor  with Rashba SOC.
%
%
%
%
%
Our analytical results show that in general the self-energy
describing the effect of an isolated impurity consists of two terms that
may have opposite signs. We find that the complete or partial cancellation of these two terms is responsible
for the presence of low-energy impurity-induced states that are not present in homogeneous SC systems~\cite{balatsky_review}.
We find that this cancellation may lead to impurity-induced subgap states even in the limit
of vanishing magnetic field. This finding does not contradict Anderson's result~\cite{AndersonT} given that in our system
the superconducting order parameter is not uniform.
%
%
For the specific case of one-dimensional (1D) heterostructures 
we study how the spectrum of the impurity-induced states changes as a function of an external magnetic field. As shown in Refs.~\cite{Sau2010, Alicea10, Lutchyn10, Oreg10}, a magnetic field may induce a quantum phase transition from a conventional (trivial) superconducting phase to a topological superconducting phase characterized by the presence of MZMs. We identify the regions in parameter space
where very low-energy impurity-induced states might affect the observation and manipulation of MZMs.
%


%

{\em Theoretical Model}.
The Hamiltonian $H$ for the heterostructure can be written as $H=H_{\rm N} + H_{\rm SC} + H_{\rm T}$,
where $H_{\rm N}$ is the Hamiltonian for the normal, i.e. non-superconducting, component (either a semiconductor or a metal),
$H_{\rm SC}$ is the Hamiltonian for the
SC and $H_{\rm T}$ is the term describing tunneling processes between the SC and the normal component.
Specifically, $H_{\rm N}$ and $H_{\rm SC}$ are defined as (henceforth $\hbar=1$)
\begin{align}
 \!H_{\rm N} &\!=\! \frac{1}{2}\sum_\kk \!\psi^\dagger_{\rm N,\kk} \left[\eps_{\rm N,\kk}\sigma_0\tau_z \!+\!
          \alpha{\bf l}_\kk\!\cdot\!\ssigma\tau_z    \!+\!
                     V_x\sigma_x\tau_z\! \right]\! \psi_{\rm N,\kk}, \label{eq:HN}\\
 \!H_{\rm SC}  &= \frac{1}{2}\sum_\kk \psi^\dagger_{\rm SC,\kk} \left[ \eps_{\rm SC,\kk} \tau_z\sigma_0 -\Delta_0 \tau_y\sigma_y \right] \psi_{\rm SC,\kk}
 \label{eq:HSC}.
\end{align}
where
$\psi^\dagger_{\kk,{\rm i}}=(c_{i,\boldsymbol{k}\uparrow}^{\dagger},\,c_{i,\boldsymbol{k}\downarrow}^{\dagger},\,c_{i,-\boldsymbol{k}\uparrow},\,c_{i,-\boldsymbol{k}\downarrow})$ is the spinor with i=$\rm N$ or i=$\rm SC$,
$c_{i,\boldsymbol{k}\sigma}^{\dagger}$ ($c_{i,\boldsymbol{k}\sigma}$) is
the creation (annihilation) operator for an electron
with momentum $\kk$ and spin $\sigma$ in the $i$-th part of the heterostructure,
$\eps_{i,\kk}= (\kk^2/2m_{i}-\mu_{i})$ with $m_i$, $\mu_i$ the electron's effective mass and chemical potential, respectively, in the i-th component,
$\sigma_j$ ($\tau_j$) are the Pauli matrices in spin (Nambu) space,
$\alpha$ is the strength of the Rashba SOC with ${\bf l}_\kk =(k_y,-k_x,0)$,
$\Delta_0$ is the amplitude of the superconducting gap,
and $V_x$ is the Zeeman splitting due to the external magnetic field along the $x$-direction.
%
%
%
The tunneling Hamiltonian can be written as
\beq
 H_T = \frac{1}{2}\sum_\kk \psi^\dagger_{\rm SC,\kk} \hat h_T(\qq) \psi_{\rm N,\kk+\qq} + h.c. \label{eq:tunneling}
\enq
where $\hat h_T(\qq)$ is the tunneling matrix. In our case, assuming that the tunneling processes conserve the spin and the momentum parallel to the SC-N interface ($\kk_\parallel$)
%
%
we have
$\hat h_T(\qq)= t \sigma_0\tau_z\delta(\qq_\parallel)$ with $t$ being the tunneling amplitude.
To quantify the effect of the tunneling term it is helpful to introduce the parameter $\Gamma_t\equiv t^2\rho_{F, {\rm SC}}$,
where $\rho_{F, {\rm SC}}$ is the density of states (DOS) of the SC at the Fermi energy, $E_{F, {\rm SC}}$.

In the presence of impurities, the Hamiltonian for the system is modified by an additional term, $H_{\rm imp}$,
describing the scattering of electrons off the impurities.
For a single isolated impurity located in the i-th ($i=\rm N, SC$) component of the heterostructure
\beq
 H_{\rm imp} = \sum_\rr \delta(\rr)\psi^\dagger_{i,\rr}\hat h_{\rm imp}\psi_{i,\rr} =
               \sum_{\kk,\kk'}  \psi^\dagger_{i,\kk}\hat h_{\rm imp}\psi_{i,\kk'}.
 \label{eq:Himp}
\enq
Here $\psi^\dagger_{i,\rr}$ ($\psi_{i\rr}$) is the creation (annihilation)
operator for an electron at position $\rr$ in the $i$-th component of the heterostructure,
and $\hat h_{\rm imp}$ is the matrix describing the structure of the impurity in spinor space.
For a scalar impurity, using the convention specified above for spinors, we have
$\hat h_{\rm imp} = \uimp \sigma_0\tau_z$ where $\uimp$ is the strength of the impurity potential.
%

The spectrum of the impurity-induced states can be obtained by locating the poles of the $T$ matrix~\cite{sm}.
Using the diagrammatic approach, one can express the $T$-matrix in terms of the Green's function for the isolated components of the heterostructure $G^{(0)}_{i}(\kk,\omega)=(\omega+i\eta - H_{i})^{-1}$ with $i=\rm N, SC$ and $\eta \rightarrow 0$.
%
%
If the impurity is located in the $i$-th component of the heterostructure, the matrix $T_i$ is given by
\beq
 T_i(\omega)=\left[\identity - \hat h_{\rm imp}\Sigma_{i,\rm imp}(\omega) \right]^{-1} \hat h_{\rm imp},
 \label{eq:Ti01}
\enq
where $\Sigma_{i,\rm imp}(\omega)=\int d\kk G_i(\kk,\omega)$ and
$G_i(\kk,\omega)$ is the Green's function of the i-th component of the heterostructure
{\em dressed} by the self-energy $\Sigma_{i,t}(\kk_\parallel,\omega)$  due to the tunneling term:
\begin{align}
 G_i(\kk,\omega) = & \left[(G^{(0)}_{i}(\kk,\omega))^{-1} - \Sigma_{i,t}(\kk,\omega), \right]^{-1}  \\
 %
 %
 \Sigma_{i,t}(\kk,\omega) = & \int d\qq \hat h_T(\qq) G^{(0)}_{\bar i}(\kk+\qq, \omega) \hat h_T(-\qq)
 \label{eq:sigma_t}.
\end{align}
Here $G^{(0)}_{\bar i}$ is the Green's function of the heterostructure's component coupled via the tunneling term to the
i-th component.
Using Eq.\eqref{eq:tunneling}, we obtain
\beq
 \Sigma_{i,t}(\kk_\parallel,\omega) = t^2 \int d\qq_\perp \sigma_0\tau_z G^{(0)}_{\bar i}(\kk_\parallel,\qq_\perp, \omega) \sigma_0\tau_z.
\enq
%
%

To understand how the presence of the tunneling term affects the spectrum of the impurity-induced states
it is useful to express $T_i$ in the following equivalent form:
\beq
 T_i=\left[\identity - \hat h_{\rm imp}\left( \Sigma^{(0)}_{i,\rm imp}(\omega) +  \Sigma^{(1)}_{i,\rm imp}(\omega) \right) \right]^{-1}\hat h_{\rm imp}
 \label{eq:Ti02}
\enq
where
\begin{align}
 \Sigma^{(0)}_{i,\rm imp}(\omega) = & \int d\kk G^{(0)}_{i}(\kk,\omega), \label{eq:sigma0} \\
 \Sigma^{(1)}_{i,\rm imp}(\omega) = & \int d\kk  G^{(0)}_{i}(\kk,\omega) \Sigma_{i,t}(\kk,\omega)_{i}G_{i}(\kk,\omega).
 \label{eq:sigma1}
\end{align}
As follows from above, there are two contributions that determine the pole structure of $T_i$:
$\Sigma^{(0)}_{i,\rm imp}$ the term that appears if the component $i$ were isolated,
and $\Sigma^{(1)}_{i,\rm imp}(\omega)$ the term due to tunneling processes between the $i$-th and $\bar i$-th component of the heterostructure. If tunneling is not a weak perturbation, the interplay between these two terms may lead to unusual properties for the spectrum of the impurity-induced states in the heterostructure.


%
\begin{figure}
\centering
\includegraphics[width=0.5\textwidth]{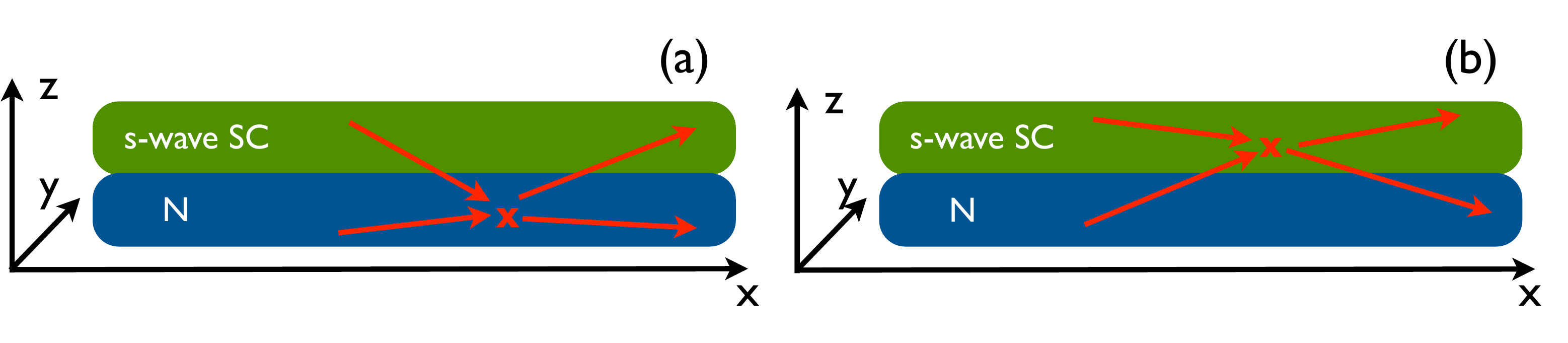}
\vspace{-0.1in}
\caption{(Color online) Sketch of 1D N/SC heterostructure with an isolated impurity, shown by the red ``X'', in the semiconductor (N), in (a), and in the SC in (b). Red arrows represent impurity-induced scattering processes.
}
\vspace{-0.2in}
\label{fig:demo}
\end{figure}

For the case when the impurity is located in the normal component (in the remainder we assume it to be a semiconductor) the effect of the tunneling term
is to induce a SC gap in it ($\Delta_{\rm ind}$) and is straightforward from Eq.~\ceq{eq:Ti01} to obtain
$T_{\rm N}(\omega) = [\tau_{z}\sigma_{0} - \uimp \hat\Sigma_{N,\rm imp}(\omega)]^{-1}\uimp$.
%
%
When no SOC is present ($\alpha=0$), $T_{\rm N}(\omega)$ does not have poles $\omega^*$ below the induced gap (i.e. $|\omega^*| \geq \Delta_{\rm ind}$). In the presence of SOC in the semiconductor the superconducting pairing will mix spin-singlet and spin-triplet pairing components, even though in the SC only s-wave pairing is present~\cite{gorkov2001,Frigeri04,triola2016}.
To further investigate this case we consider the quasi-1D system shown in Fig.~\ref{fig:demo}
in which $L_x\to\infty$ and $L_y$, $L_z$ are small enough so that the spectrum is comprised of 1D subbands, $\eps_{k_x}^{(n)}$,
with energy separation larger than $\Delta_0$.
%
For concreteness in the remainder we limit ourselves to the case in which only one spinful subband is occupied.
When $V_x$ is larger than a critical value, $V_x^{c}$, the system
is expected to be in a topological phase~\cite{Lutchyn10, Oreg10}. 
%
%
%
\begin{figure}[t]
\centering
\includegraphics[width=2.8in,clip]{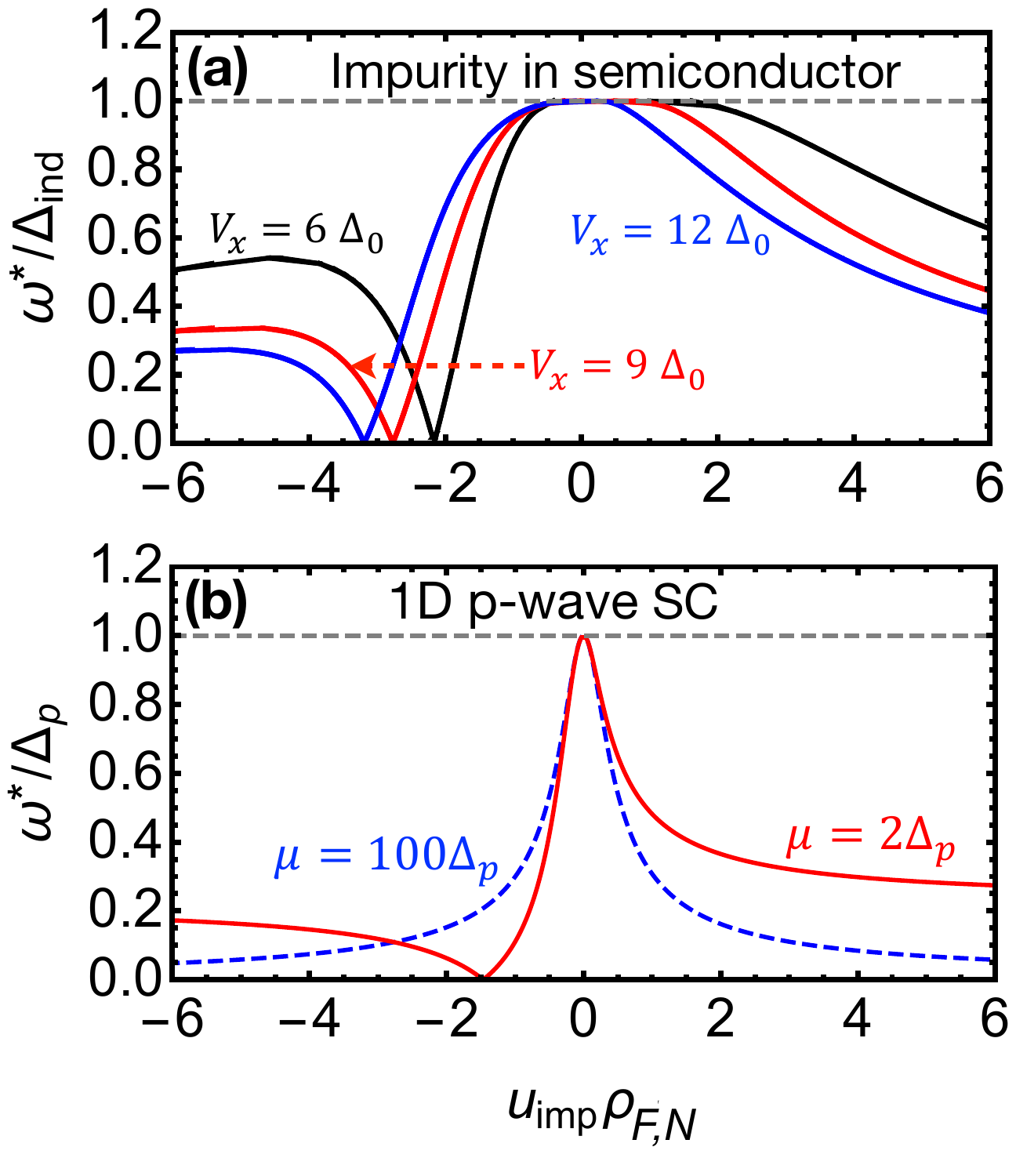}
\vspace{-0.2in}
\caption{
         (Color online)
         (a) Spectrum of impurity-induced bound states for 1D N/SC heterostructure as a function of $\uimp\rho_{F,N}$
         for the case (a) of Fig.\ref{fig:demo} and $V_x>V_x^c$.
         Here $k_{F,\rm \rm N}^2/(2m_N)=1.5 \Delta_0$, $\alpha_{SO} k_{F,N}=4.2 \Delta_0$, $\mu_N=1.5
         \Delta_0$, $\Gamma_t=5\Delta_0$, $V_x^c \approx  5.2\Delta_0$.
         (b) Spectrum of impurity-induced bound states for a 1D p-wave SC as a function  $\uimp\rho_{F,N}$ for different values of $\mu$.
}
\label{fig:IBS_SM_pwave}
\end{figure}
For parameter values relevant for current experiments~\cite{sm} for $V_x<V_x^c$ the impurity-induced states have energies, $\omega^*$, very close to
the induced-gap edge. When the chemical potential is much larger than the SC bulk gap~\cite{sm}, in the trivial regime,  $|\omega^*|$ can be smaller
than $\Delta_{\rm ind}$, albeit it does not approaches zero.
The spectrum of the impurity-induced states
is completely different in the topological regime. In this regime the induced superconducting pairing is p-wave and we find that the energy of the bound states:
(i)    depends very strongly on $\uimp$,
(ii)   it is strongly asymmetric with respect to $\uimp=0$,
(iii)  it can go to zero for finite (negative) values of $\uimp$~\cite{SauDemler13}.
This can be seen in
Fig.~\ref{fig:IBS_SM_pwave}~(a) where the dependence of $\omega^*$ on $\uimp\rho_{F,N}$  ($\rho_{F,N}$ being the DOS of the semiconductor (N) at its Fermi energy $E_{F,N}$) for different values of the Zeeman splitting $V_x>V_x^{c}$.
%
%
%
%

The results shown in Fig.~\ref{fig:IBS_SM_pwave} (a) can be qualitatively understood considering a scalar impurity, Eq.~\ceq{eq:Himp},
in a 1D p-wave superconductor for which:
$H_{\rm pSC}=\sum_{k_x\sigma\sigma'}[c^\dagger_{k_x,\sigma}(k_x^2/2m-\mu)\sigma_0 c_{k_x,\sigma'} +
                                    i \Delta_{p}(k_x/k_F) c^\dagger_{k_x,\sigma}\dd_{k_x}\cdot{\ssigma}\sigma_yc^\dagger_{-k_x,\sigma'} + {\rm h.c.}]$,
where $\Delta_{p}(k_x/k_F)=-\Delta_{p}(-k_x/k_F)$ is the amplitude of the superconducting p-wave pairing and $\dd_{k_x}$ is the
unit vector characterizing the polarization of the triplet state~\cite{mackenzie2003}.
In this case $T(\omega)=\uimp[\tau_z -\uimp\int d k_x G_{\rm p-SC}(\omega, k_x)]^{-1}$,
where $G_{\rm p-SC}(\omega, k_x)=(\omega+i\eta-H_{\rm p-SC})^{-1}$.
Due to the 1D nature of the carriers, one finds that, at low energies, the
density of states is strongly dependent on their energy $\epsilon$: $\rho(\epsilon)~\approx 1/\sqrt{\epsilon}$.
This fact makes the energy of the impurity bound state strongly dependent on $\uimp$ when
$\mu_N$ is close to the bottom of the band. This is shown in Fig.~\ref{fig:IBS_SM_pwave} (b)
where we can see that the energy of the bound state depends
strongly on $\uimp$ when $\mu$ is small (solid line) and fairly weakly for large $\mu$ (dashed line)~\footnote{A similar asymmetry for the energy of bound states with respect to $\uimp=0$ has also been found
for the case of d-wave superconductors~\cite{Atkinson00}: also in that case such asymmetry is
due to the dependence of the DOS on the energy.}.
We should emphasize that this asymmetry effect is very relevant for 1D topological SC wires supporting MZMs in which typically $\mu_N$
must be quite small, i.e. $|\mu_N| < \sqrt{V_{x}^2-\Delta_{\rm ind}^2}$~\cite{Sau2010, Alicea10, Lutchyn10, Oreg10}.
%
%

In the most recent realizations of 1D topological SC wires~\cite{lutchynreview2017, Albrecht16, Zhang16, Deng2016}
the semiconductor and the interface between the semiconductor and the SC are of very high quality so that very few impurities
are expected to be present in the semiconductor or at the interface. On the other hand, the SC (i.e. aluminum) is disordered. Therefore, henceforth we consider
the situation in which the impurities are located in the SC.
In this case, using Eq.~\ceq{eq:Ti02} one finds
\begin{equation}
 T_{\rm SC}=\frac{\uimp}{\tau_{z}\sigma_{0}-\uimp \Sigma_{SC,\rm imp}^{(0)}(\omega) - \uimp \Sigma_{SC,\rm imp}^{(1)}(\omega)}.
 \label{eq:Tmatrix}
\end{equation}
For the case in which the SC is s-wave and the tunneling is such that $\hat h_T=t\delta(\qq_\parallel)\sigma_z\tau_0$
we obtain
\begin{align}
 \Sigma_{SC,\rm imp}^{(0)}(\omega) =& -\frac{\rho_{F,SC}}{\sqrt{\Delta_0^2-\omega^2}}\left[\omega\sigma_0\tau_0 + \Delta_0\sigma_y\tau_y \right] \\
 \Sigma_{SC,\rm imp}^{(1)}(\omega) =& \int d\kk_\parallel\int d \kk_\perp G^{(0)}_{SC}(\kk_\parallel,\kk_\perp,\omega) \nonumber \\
                               &  \Sigma_{SC,t}(\kk_\parallel,\omega)
                                  G_{SC}(\kk_\parallel,\kk_\perp,\omega)
\end{align}
with $\Sigma_{SC,t}(\kk_\parallel,\omega)$ given by Eq.~\ceq{eq:sigma_t}.
%
%
One can show that the strength of the second term $\Sigma_{SC,\rm imp}^{(1)}(\omega)$
is proportional to the dimensionless parameter
$\alpha_{SwS}=\frac{\Gamma_{t}}{E_{\rm F,N}}\frac{k_{F, {\rm N}}}{k_{F, {\rm SC}}}$~(see \cite{sm} for details)
where $k_{F, {\rm N}}$, $k_{F, {\rm SC}}$ are the Fermi momenta in the N and SC, respectively.
%
%
%
%
%

\begin{figure}[t]
\centering
\vspace{0.15in}
\includegraphics[width=3.4in,clip]{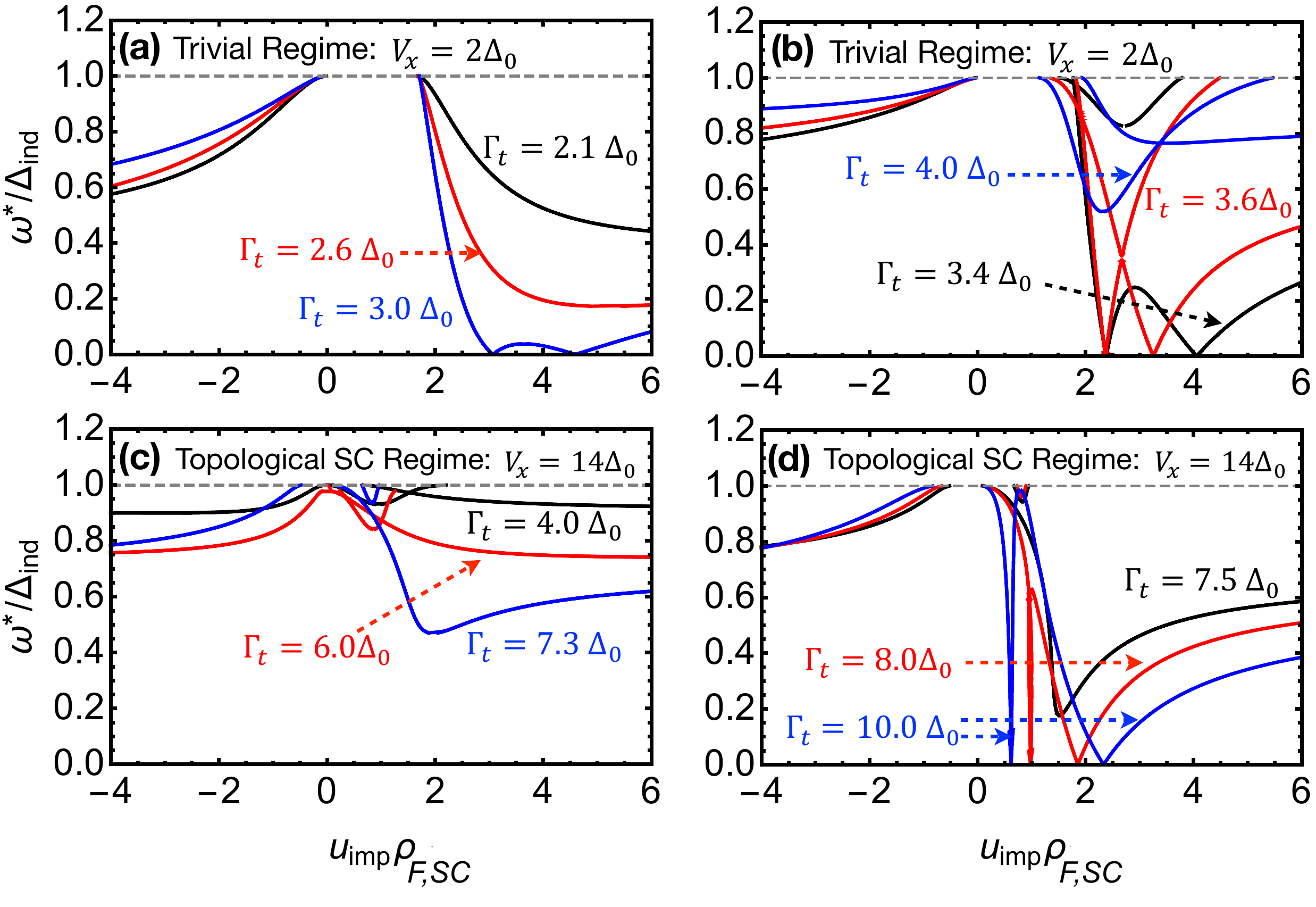}
\vspace{-0.2in}
\caption{
         (Color online)
         Spectrum of impurity-induced bound states as function of $\uimp\rho_{F,SC}$ for 1D N/SC heterostructure when the impurity is located in the SC
         in the trivial regime, (a) and (b), and topological regime (c) and (d).
         Here $k_{F,\rm N}^2/(2 m_N)=1.5\Delta_0$, $\mu_N=1.5\Delta_0$, $\alpha_{SO}k_{F, \rm N}=4.2 \Delta_0$, $k_{F,\rm N}/k_{F,\rm SC}=0.3$.
}
\vspace{-0.2in}
\label{fig:Interf_TopoSC_Trivial}
\end{figure}

Fig.~\ref{fig:Interf_TopoSC_Trivial}~(a) shows the spectrum of the impurity-induced states as a function of $\uimp\rho_{F,SC}$
for the 1D case in which the N/SC heterostructure is in the topologically trivial phase, $V_x=2\Delta_0<V_x^{(c)}$,
and different values of $\Gamma_t$.
In the limit $\alpha_{SwS} \rightarrow 0$,
$t\neq 0$ (i.e. $\Sigma_{SC,\rm imp}^{(1)}\to 0$ and $\Delta_{\rm ind}\neq 0$), we find bound states close to the gap edge. As $\alpha_{SwS}$ increases, the interplay between $\Sigma_{SC,\rm imp}^{(0)}$ and $\Sigma_{SC,\rm imp}^{(1)}$ may lead to low-lying subgap states as demonstrated in Figs.~\ref{fig:Interf_TopoSC_Trivial} (a) and (b).
The results of Fig.~\ref{fig:Interf_TopoSC_Trivial}~(b) also show that as $\Gamma_t$ increases the spectrum
of the impurity-induced bound states becomes more asymmetric with respect to $\uimp=0$ as we have found
for the case in which the impurity is located in the N.
It is very interesting to notice that, contrary to the case when an impurity is located in the N,
see Fig~\ref{fig:SI_IBS_Lmu}~(a) in \cite{sm},
an impurity in the SC may lead to low-lying subgap
states with $\omega^*\rightarrow 0$ in the trivial regime.

Figs.~\ref{fig:Interf_TopoSC_Trivial}~(c),~(d) show the results when the N/SC heterostructure is in the topological phase
$V_x=14\Delta_0>V_x^{(c)}$. One can see that the spectrum is strongly asymmetric in this case even for relatively small values of
$\Gamma_t$, Figs.~\ref{fig:Interf_TopoSC_Trivial}~(c).
For larger $\Gamma_t$ we find that also in the topological phase the impurity can induce zero energy bound states for
relatively small values of $\uimp\rho_{F,SC}$, Figs.~\ref{fig:Interf_TopoSC_Trivial}~(d). These results suggest
that in the topological phase the value of $\uimp$ necessary to induce a zero-energy bound state decreases as $\Gamma_t$ increases. Thus, there is an optimal value of $\Gamma_t$ for which the induced gap is large and, at the same time,  impurities in the SC do not result in significant subgap density of states.
%

%
%
%

\begin{figure}[h]
\centering
\vspace{0.15in}
\includegraphics[width=3.4in,clip]{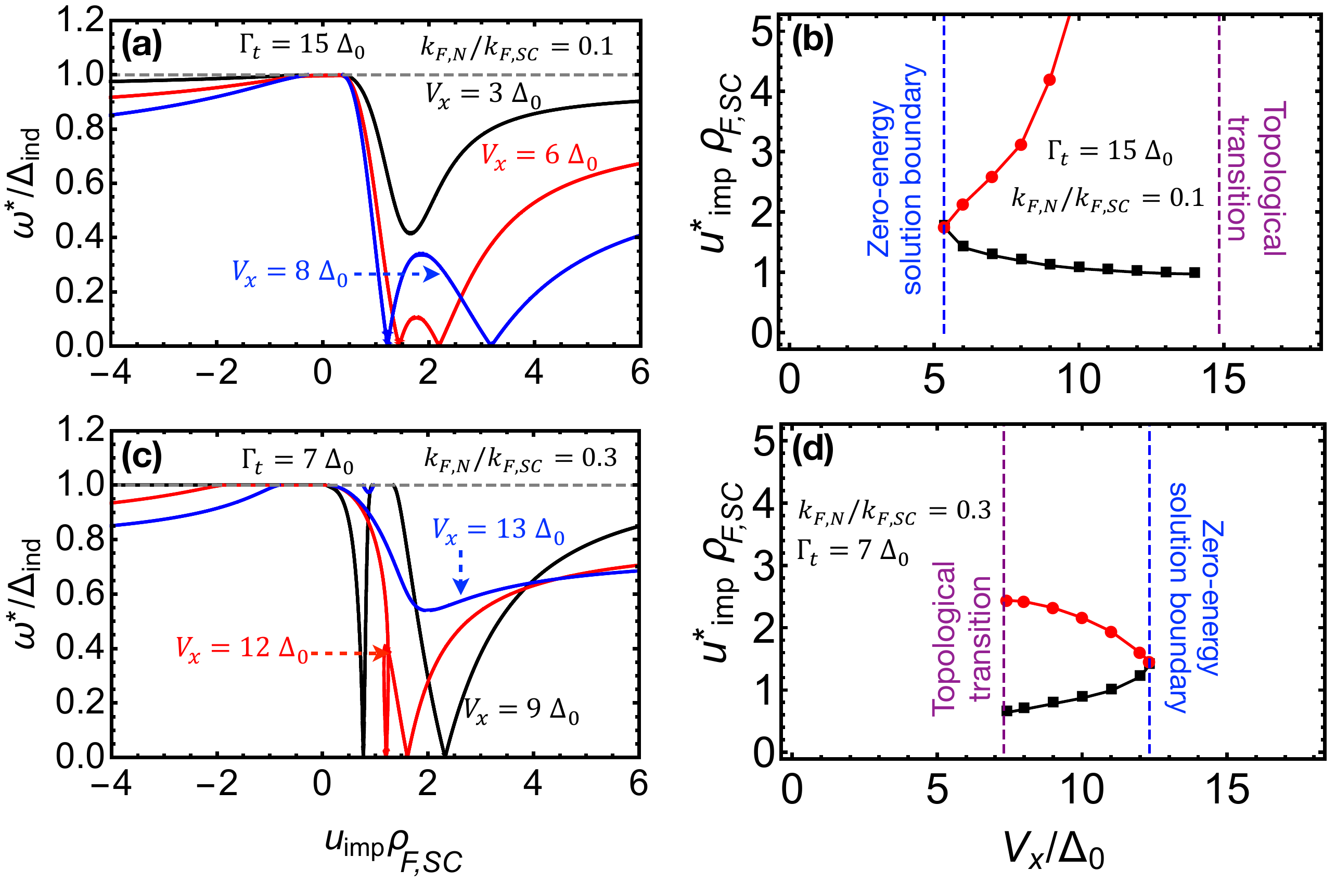}
\vspace{-0.2in}
\caption{
(Color online)
Evolution of impurity-induced bound states spectrum with Zeeman field. 
(a),~(b):  trivial regime,
$\Gamma_t=15\Delta_0$, $k_{F,\rm N}/k_{F,\rm SC}=0.1$. 
(c),~(d): topological regime, $\Gamma_t=7\Delta_0$, $k_{F,\rm N}/k_{F,\rm SC}=0.3$.
%
}
\label{fig:Interf_Zeeman}
\end{figure}

The spectrum of the impurity bound states as a function of Zeeman coupling for $V_x<V_x^{(c)}$ and
fixed $\Gamma_t$ is shown in Fig.~\ref{fig:Interf_Zeeman}~(a). As one can see there is a threshold value of $V_x$ for the emergence of bound states with $\omega^*\rightarrow 0$. We plot the value of $|u^*_{imp}\rho_{F,SC}|$ such that $\omega^*=0$ as a function of $V_x$ in Fig.~\ref{fig:Interf_Zeeman}~(b): one of the solutions decreases and approaches a constant at the topological transition whereas the other one increases to infinity. Similarly, we study the topological SC regime in Figs.~\ref{fig:Interf_Zeeman}~(c), ~(d). As we increase $V_x$, two zero-energy solutions merge and then disappear.

%
%
%
%

\begin{figure}[h]
\centering
\vspace{0.15in}
\includegraphics[width=3.4in,clip]{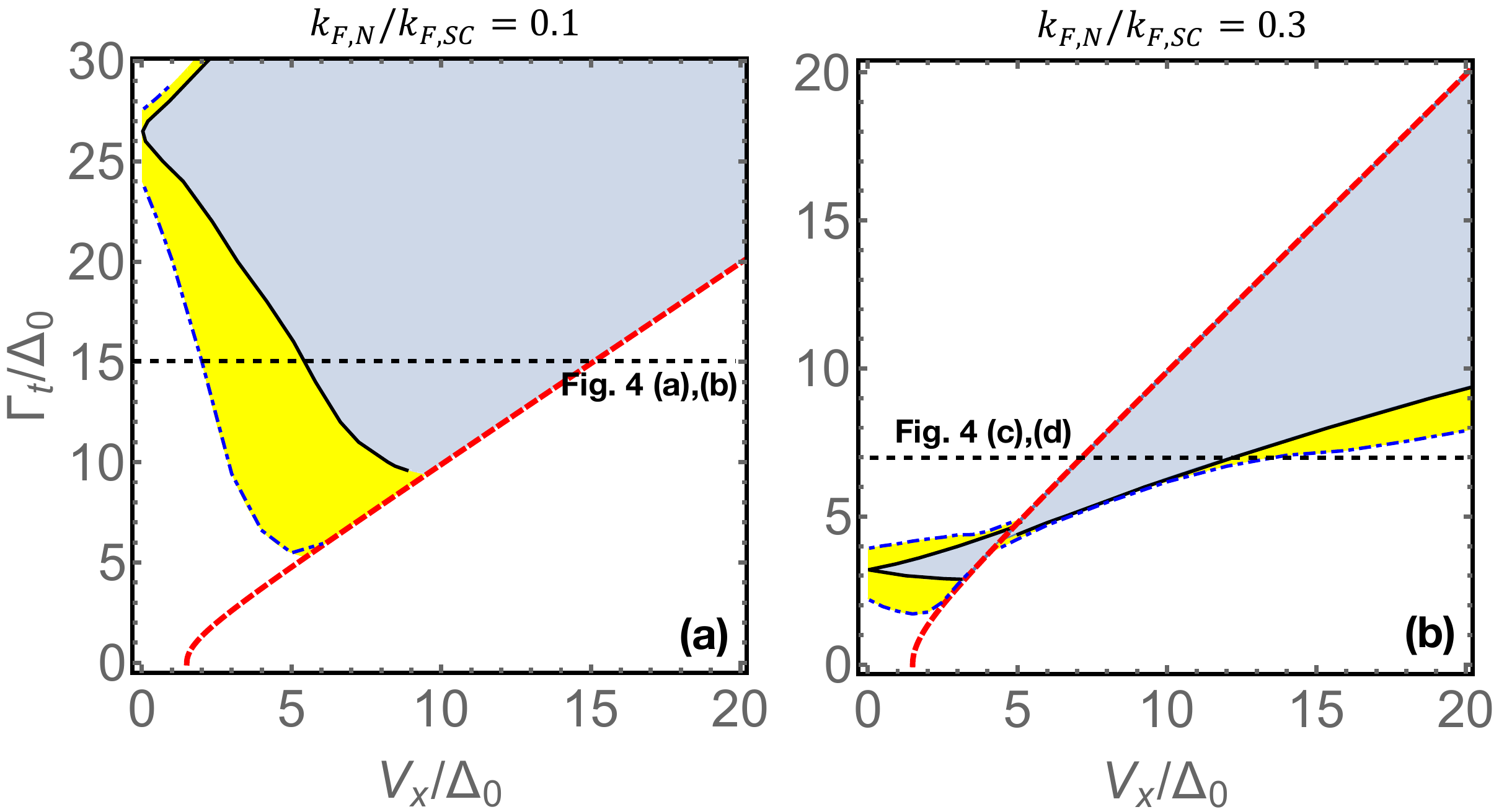}
\vspace{-0.2in}
\caption{
         (Color online)
         Phase diagram in  $(V_x,\Gamma_t)$ plane identifying the regions, shown in grey (yellow)
         for which $\omega^*=0$ ($\omega^*<0.6 \Delta_{\rm ind}$) for some
         finite value of $\uimp$.
         The red dashed line shows the boundary between trivial and topological regime.
         Horizontal dashed line is placed at the value of $\Gamma_t$ for which the results of Fig.~\ref{fig:Interf_Zeeman} were obtained.
         Here $k_{F,\rm N}^2/(2 m_N)=1.5\Delta_0$, $\mu_N=1.5\Delta_0$, $\alpha_{SO}k_{F, \rm N}=4.2 \Delta_0$.
         }
%
\label{fig:PhaseD_Gamma}
\end{figure}

%
Considering that $V_x$ and $\Gamma_t$ are two of the key parameters that can be controlled in experiments to realize MZMs
in proximitized nanowires, the knowledge of where in  the $(V_x,\Gamma_t)$ plane $\omega^*=0$ is of great importance for the realization of topological qubits based on such systems~\cite{Hyart13, aasen2016, Plugge16b, Karzig16}.
Figs.~\ref{fig:PhaseD_Gamma}~(a),~(b) show in grey-blue (yellow) the regions in the $(V_x,\Gamma_t)$ plane for which there exist a finite
value of $\uimp$
such that $\omega^*=0$ ($\omega^*<0.6 \Delta_{\rm ind}$). 
The red dashed line shows the boundary between trivial and topological regimes.
The horizontal dashed line 
in Fig.~\ref{fig:PhaseD_Gamma}~(a)
(Fig.~\ref{fig:PhaseD_Gamma}~(b))
identifies the valuesof $\Gamma_t$ for which the results of Fig.~\ref{fig:Interf_Zeeman}~(a),~(b)
(Fig.~\ref{fig:Interf_Zeeman}~(c),~(d))
were obtained.
As follows from Fig.~\ref{fig:PhaseD_Gamma}~(a), the area where $\omega^*=0$ is rather large in the trivial regime and becomes smaller in the topological one when $k_{F, \rm N}/k_{F, \rm SC} \ll 1$.
As the ratio  $k_{F, \rm N}/k_{F, \rm SC}$ increases the area where $\omega^*=0$ decreases in the trivial phase and increases in the topological phase,
as shown by Fig.~\ref{fig:PhaseD_Gamma}~(b).
Thus, the ratio of $k_{F, \rm N}/k_{F, \rm SC}$ is an important parameter when trying to reduce disorder effects in superconducting heterostructures.
For aluminum-based proximitized nanowires this parameter is quite small, $k_{F, \rm N}/k_{F, \rm SC} \ll 1$. The parameter $\alpha_{SwS}$ can be controlled experimentally by changing the back-gate voltage in proximitized nanowires~\cite{Sole2017} so the propensity for the formation of impurity-induced bound states we predict, see Fig.~\ref{fig:PhaseD_Gamma}, can be tested experimentally.
%

%

{\em Conclusions.}
%
%
We have studied impurity-induced subgap states in superconductor-based heterostructures. In the case of proximitized nanowires, considered in this work in detail, we find that in these structures there is a large region in parameter space for which the impurities in the superconductor can induce low energy states even when the superconductor is purely s-wave.
Our work presents results for the spectrum of the bound states induced by a single impurity and so is complementary
to the previous studies that considered the case of
many weak impurities~\cite{Potter2011, Potter2011a, Lobos2012, lutchyn2012, Sau2012, Tkachov2013, Sau2013, Hui2015, Cole2016, Liu2017} via disorder-averaging techniques.
Our results are directly relevant to experimental situations in which the impurity density is low and
disorder-averaging is not justified.
In addition, they are instrumental to extend the study of the effect of many-impurities via disorder-averaging
to the unitary limit, i.e. the limit of strong impurities,
both for the case when the impurities are located in the semiconductor and the case when they are located in the superconductor.
%

Our results provide guidance for the optimization of superconductor-semiconductor heterostructures: although a strong tunneling is beneficial to obtain a large gap it also enhances the effect of the impurities located in the s-wave superconductor on the superconducting state induced in the semiconductor. Therefore, we find that, when the effect of impurities is included, the optimal coupling to the superconductor is not strong but intermediate, i.e. $\Gamma_t \sim \Delta_0$.


{\it Acknowledgments.}
ER acknowledges support from NSF CAREER DMR-1455233, ARO-W911NF-16- 1-0387, and ONR-N00014-16-1-3158.

\appendix

\begin{widetext}

\section{Supplementary Information for ``Impurity-induced states in superconducting heterostructures''}

\begin{figure}[b]
\centering
\includegraphics[width=6in,clip]{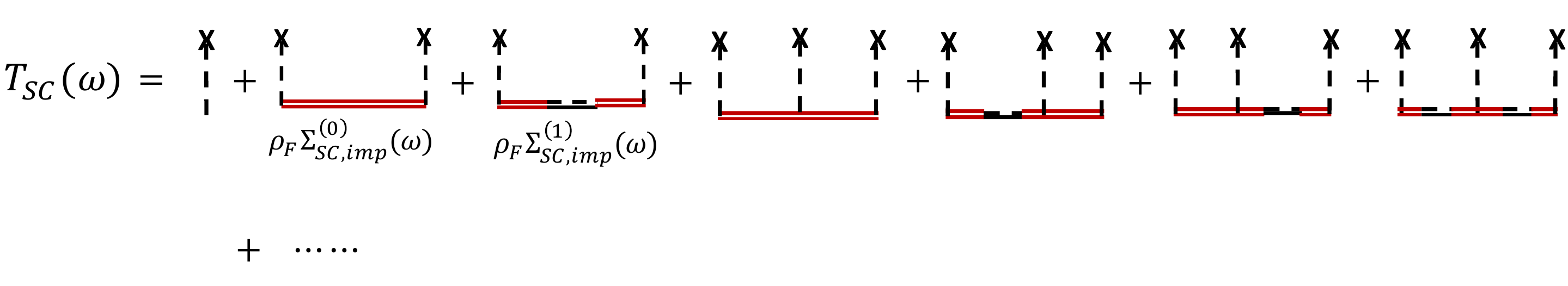}
\vspace{-0.1in}
\caption{Diagrammatic representation of the scattering T-matrix for a single impurity in the superconductor of a superconducting heterostructure.
Processes that involve two scatterings can be of two types: 1) processes in which the electrons do not leave the SC between the two scatterings,
2) processes in which the electrons travel through the N in between the two scatterings.
The meaning of the different lines is the following:
double solid line (red): propagator in s-wave SC;
dashed-solid line (black): propagator in semiconductor wire with proximity induced superconducting gap;
cross head - dashed line: scattering off the impurity.
}
\label{fig:SI_Tmatrix_FD}
\vspace{0.15in}
\end{figure}

In this supplementary material we provide: (i) details on the derivation of the T-matrix expression
for the case when the impurity is located in the superconductor, see Eq. (13-15) of the main text,
(ii) the relation between the parameters values used in our calculations and the parameters values of current experiments on quasi 1D SM-SC heterostructures,
(iii) the spectrum of the impurity-induced states for the case when the impurity is located in the N and the chemical potential in the N
is much larger than the SC's gap.

\section{T-matrix calculation for an impurity in the superconductor}
The scattering T-matrix for a single impurity in a superconductor proximity-coupled to a semiconductor nanowire can be described by a diagrammatic representation shown in Fig. \ref{fig:SI_Tmatrix_FD}:
\begin{align}
T_{\rm SC}(\omega)&= u_{imp} \tau_{z} \sigma_{0} + u_{imp}^2 \tau_{z} \sigma_{0}\cdot \left(\Sigma_{SC,imp}^{(0)}(\omega)+\Sigma_{{\rm SC}, imp}^{(1)}(\omega)\right)\cdot \tau_{z} \sigma_{0}\nonumber\\
&\quad +u_{imp}^3 \tau_{z} \sigma_{0}\cdot \left(\Sigma_{SC,imp}^{(0)}(\omega)+\Sigma_{SC,imp}^{(1)}(\omega)\right)\cdot \tau_{z} \sigma_{0} \cdot \left(\Sigma_{SC,imp}^{(0)}(\omega)+\Sigma_{SC,imp}^{(1)}(\omega)\right)\cdot \tau_{z} \sigma_{0} + \cdots \nonumber\\
&= \frac{u_{imp}}{\tau_{z} \sigma_{0}-u_{imp}\Sigma_{SC,imp}^{(0)}(\omega) - u_{imp}\Sigma_{SC,imp}^{(1)}(\omega)},
\label{eq:SI_TmatrixD}
\end{align}
where $\Sigma_{SC,imp}^{(0)}(\omega)$ represents the contribution to the self-energy for a clean s-wave superconductor:
\begin{equation}
\Sigma_{SC,imp}^{(0)}(\omega)=\sum_{\overrightarrow{k}}G_{SC}^{(0)}(\omega,\overrightarrow{k})=\rho_{F,SC}\,g^{qc}(\omega)
=-\frac{\rho_{F,SC}}{\sqrt{\Delta_0^{2}-\omega^{2}}}
\left(\begin{array}{cc}
\omega \sigma_0 & (\Delta_0 i\sigma_y)^{\dagger}\\
 (\Delta_0 i\sigma_y) & \omega \sigma_0
\end{array}\right).
\label{eq:SI_self_dis_S}
\end{equation}
Above expression corresponds to Eq. (14) of the main text. The second term $\Sigma_{SC,imp}^{(1)}(\omega)$ represents a process of an electron tunneling between the SC and semiconductor nanowire and scattering off the impurity
\begin{align}
\Sigma_{SC,imp}^{(1)}(\omega) &= \int d\kk_\parallel d\kk_{1,\perp} d\kk_{2,\perp} G^{(0)}_{SC}(\kk_\parallel,\kk_{1,\perp} ,\omega)\Sigma_{SC,t}(\kk_\parallel,\omega) G_{SC}(\kk_\parallel,\kk_{2,\perp} ,\omega),  \label{eq:GimpAA}.\\
& =t^2  \int d\kk_\parallel d\kk_{1,\perp} d\kk_{2,\perp} G^{(0)}_{SC}(\kk_\parallel,\kk_{1,\perp},\omega) \cdot \tau_{z} \sigma_{0}  \cdot G_{N}(\kk_\parallel,\omega) \cdot \tau_{z} \sigma_{0}  \cdot G^{(0)}_{SC}(\kk_\parallel,\kk_{2,\perp},\omega) \label{eq:Gimp1}.
\end{align}
Here $t$ is tunneling matrix element between the SC and nanowire (N), $G_{N}$ is the dressed semiconductor Green function (in proximity to a clean SC). We assume that momentum parallel to the SC-N interface is conserved.

The largest contribution to the scattering in SC comes from on-shell processes (i.e. close to the Fermi surface). Therefore, it's convenient to introduce $\delta \kk\equiv (\delta k,\hat{\Omega})$, where $\delta k=|\kk-\kk_{F, \rm SC}| \ll k_{F,SC}$ and $\hat{\Omega}\equiv \kk/|\kk|$.
One can approximate the quasiparticle energy spectrum in the superconductor $\epsilon(k)$ as $\epsilon(k)= v_{F, \rm SC}  \delta k$ where $v_{F, \rm SC}$ is the Fermi velocity in the superconductor. After integration over $k$, the bulk Green's function in the SC, $G^{(0)}_{SC}$,  becomes almost independent of the momentum $\hat{\Omega}$. Since the integral over $k$ mostly comes from the contribution of pole near Fermi energy, one can perform the integration over $k$ analytically:
\begin{align}
\Sigma_{SC,imp}^{(1)}(\omega) \!=\!\frac{L_{z}k_{F,N}}{E_{F,N}}t^{2}\!\! \int \!d\hat{\Omega}\!\left(\!\frac{\rho_{F,SC}}{k_{F,SC}L_{z}}\int d \delta k G_{SC}(\omega; \delta k,\hat{\Omega})\right)\cdot\tau_{z}\sigma_{0}\cdot R(\omega)\cdot\tau_{z}\sigma_{0}\cdot \int d\hat{\Omega}\!\left(\!\frac{\rho_{F,SC}}{k_{F,SC}L_{z}}\int d\delta k G_{SC}(\omega; \delta k,\hat{\Omega})\!\right),
\end{align}
where $k_{F,N}$ and $E_{F,N}$ are the Fermi wavevector and Fermi energy in the semiconductor wire, respectively,
$\rho_{F,SC}$ the normal-state density of states (DOS), at the Fermi energy, of the SC,
and
\begin{equation}
R(\omega)\equiv E_{F,N}\int\frac{d\widetilde{\kk}_\parallel}{2\pi}G_{N}(\omega;\widetilde{\kk}_\parallel)\;\text{with $\widetilde{\kk}_\parallel=\frac{\kk_\parallel}{k_{F,N}}$}.
\end{equation}
Notice that the DOS contributing to the N-SC tunneling amplitude is given by $\rho_{F,SC}^{t} = \rho_{F,SC}/(k_{F,SC}L_{z})$).
%
Assuming the bare SC Green's function $G_{SC}(\omega;k,\hat{\Omega})$ to be isotropic and, thus, independent of $\hat{\Omega}$, one can simplify the expression for  $\Sigma_{SC,imp}^{(1)}(\omega)$:
\begin{equation}
\Sigma_{SC,imp}^{(1)}(\omega)=\alpha_{SwS} \rho_{F,SC}\left[g^{qc}(\omega)\cdot\tau_{z} \sigma_{0}\cdot R(\omega)\cdot\tau_{z} \sigma_{0}\cdot g^{qc}(\omega)\right]
\label{eq:self_dis_SwS}
\end{equation}
where the dimensionless parameter $\alpha_{SwS}$ reads
\begin{align}
\alpha_{SwS}&=\frac{\Gamma_{t}}{E_{F,N}}\frac{k_{F,N}}{k_{F,SC}} \; \text{where $\Gamma_{t}=\pi |t|^2\rho_{F,SC}^{t}$} \label{eq:alphaSwS}.
\end{align}
Notice that the presence of the vertex matrix $\tau_{z} \sigma_{0}$ flips the position of the zero energy solutions from $u_{\rm imp}^*\rho_{F,SC}<0$ to $u_{\rm imp}^*\rho_{F,SC}>0$ (please compare Fig. 2 with Fig. 3 and 4 in the main text).

\begin{figure}[t]
\centering
\includegraphics[width=4.8in,clip]{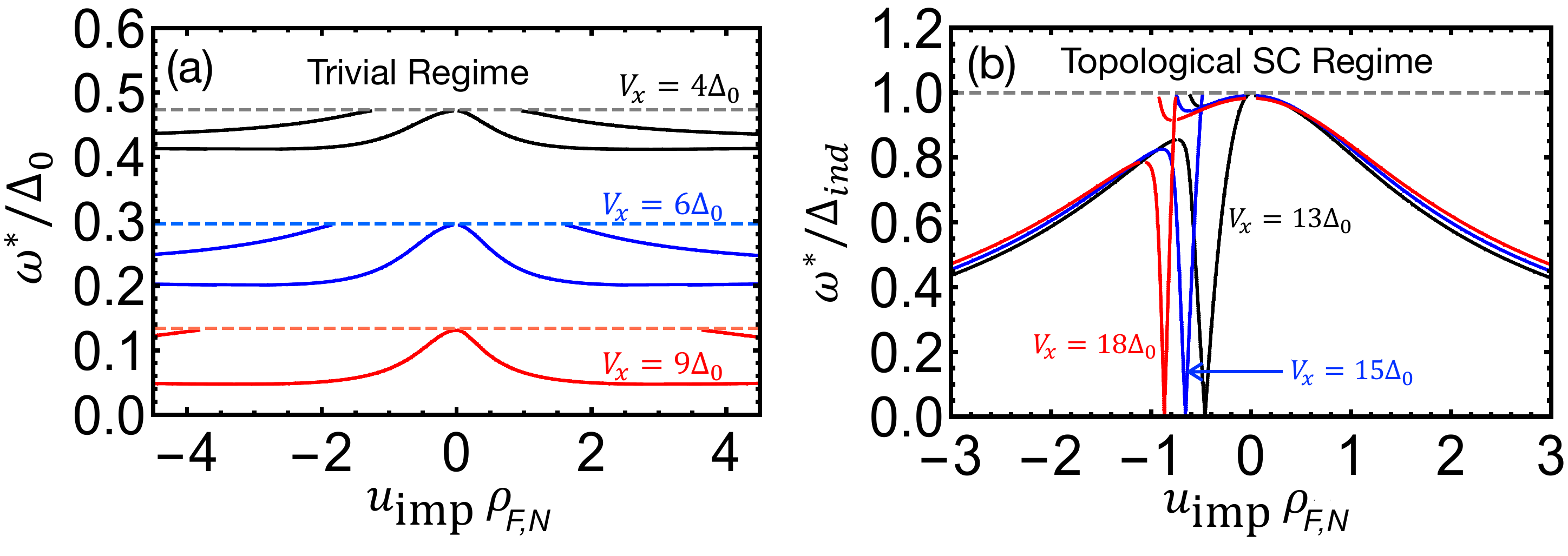}
\vspace{-0.2in}
\caption{Spectrum of impurity-induced bound states for 1D N/SC heterostructure as a function of $\uimp\rho_{F,N}$
         for the case in which the impurity is in the semiconductor. $k_{F,\rm \rm N}^2/(2m_{w})=10 \Delta_0$, $\alpha k_F=3 \Delta_0$, $\Gamma_t=5\Delta_0$
         in the trivial regime, (a), and topological regime (b). Here the topological transition occurs at $V_x^c \approx 11.2\Delta_0$
}
\label{fig:SI_IBS_Lmu}
\end{figure}

\section{Parameters used in the calculation vs Experimental values}

Here, we briefly explain how to choose the numerical parameters based on relevant experimental systems \cite{Beenakker13a,Alicea12a,Leijnse12,Stanescu13b, lutchynreview2017}.
We consider aluminum/InSb (SC/N), and choose for the superconducting gap of the bulk SC $\Delta_0=0.2$~meV, effective mass of the semiconductor (N) $m_{\rm eff}=0.014 m_{\rm e}$ with $m_{\rm e}$  the electron's mass, and the Rashba spin-orbit coupling strength of semiconductor $\alpha_{SO} = 0.2 - 1 {\rm eV} \cdot \ang$. We consider the energy dispersion of semiconductor wire (without Rashba spin-orbit coupling and Zeeman splitting) as $\epsilon_{N,k}=\frac{k^2}{2 m_{\rm eff}}-\mu_N=\frac{k_{F,N}^2}{2 m_{\rm eff}}(\hat{k}^2-1)$
with $\hat{k}=k/k_{F,N}$. Including both Rashba coupling and Zeeman splitting, the energy spectrum $E_{N}(\hat{k})=\frac{k_{F,N}^2}{2 m_{\rm eff}}(\hat{k}^2-1)\pm \sqrt{V_x^2+(\hat{k} k_{F,N} \alpha_{SO})^2}$ with the Fermi surface corresponding to $E_{N}(\hat{k}^*)=0$, and so the Rashba spin-orbit energy can be written as $E_{SO}=\hat{k}^* k_{F,N} \alpha_{SO}$.
In the numerical calculation of the main text, we used the parameters $\Delta_0=0.2$~meV, $m_{\rm eff}=0.014 m_{\rm e}$,
$\alpha_{SO} = 0.8 {\rm eV} \cdot \ang$. We choose $\frac{k_{F,N}^2}{2 m_{\rm eff}}=1.5\Delta_0$, from which we get $k_{F,N}\alpha_{SO}=4.22\Delta_0$.

\section{Impurity-induced bound states in the topologically trivial regime $V_x < V_x^{c}$}

In this section, we consider the case in which the impurity in the semiconductor, and show that low energy bound states appear even in the topological trivial regime, if the chemical potential is large.
In this section we assume $\Delta_0=0.2$~meV, $m_{\rm eff}=0.014 m_{\rm e}$, $\alpha_{SO} = 0.22 {\rm eV} \cdot \ang$, and $\frac{k_{F,N}^2}{2 m_{\rm eff}}=10.0 \Delta_0$, and therefore $k_{F,N}\alpha_{SO}=3.0\Delta_0$.

rather than those with smaller chemical potential in the main text, we choose a different parameter set:  $\Delta_0=0.2 meV$, $m_{\rm eff}=0.014 m_{\rm e}$, $\alpha_{SO} = 0.22 eV \cdot \ang$. We choose $\frac{k_{F,N}^2}{2 m_{\rm eff}}=10.0 \Delta_0$, from which we get $k_{F,N}\alpha_{SO}=3.0\Delta_0$.
Fig.~\ref{fig:SI_IBS_Lmu}~(a) and (b) show how $\omega^*$ depends on
$\uimp\rho_{F,N}$
for different values of the Zeeman splitting $V_x$.
Fig.~\ref{fig:SI_IBS_Lmu}~(a) (Fig.~\ref{fig:SI_IBS_Lmu}~(b)) shows the results for $V_x<V_x^{c}$ ($V_x>V_x^{c}$),  the dashed lines show the value of $\Delta_{\rm ind}$. Interestingly, we can see that impurity bound states (albeit with non-zero energy) can appear even in the topologically trivial regime.
We also notice that the bound states shift to the induced-gap edge if we decrease the parameter $\frac{k_{F,N}^2}{2 m_{\rm eff}}$,.
We see that in the trivial regime the energy of the impurity-induced states becomes smaller as $V_x$ increases but it never goes to zero. We also notice that the energy of the bound states depends weakly on $u_{\rm imp}$ with a slight asymmetry of the spectrum with respect to the sign of the potential of the impurity.

\end{widetext}

%

\end{document}